\documentclass[a4paper,11pt]{article}
\usepackage{pos}
\usepackage{graphicx}
\usepackage{lineno}

\usepackage{amsmath}	% Advanced maths commands
\usepackage{amssymb}	% Extra maths symbols

{}

\title{Baikal-GVD Astrophysical Neutrino Candidate near the Blazar TXS~0506+056}
\author[1]{V.M.~Aynutdinov}
\author[2]{V.A.~Allakhverdyan}
\author[1]{A.D.~Avrorin}
\author[1]{A.V.~Avrorin}
\author[3,4]{Z.~Barda\v{c}ov\'{a}}
\author[2]{I.A.~Belolaptikov}
\author[1]{E.A.~Bondarev}
\author[2]{I.V.~Borina}
\author[5]{N.M.~Budnev}
\author[12]{V.A.~Chadymov}
\author[6]{A.S.~Chepurnov}
\author*[2,7]{V.Y.~Dik}
\author[1]{G.V.~Domogatsky}
\author[1]{A.A.~Doroshenko}
\author[3]{R.~Dvornick\'{y}}
\author[5]{A.N.~Dyachok}
\author[1]{Zh.-A.M.~Dzhilkibaev}
\author[3,4]{E.~Eckerov\'{a}}
\author[2]{T.V.~Elzhov}
\author[4]{L.~Fajt}
\author[12]{V.N. Fomin}
\author[5]{A.R.~Gafarov}
\author[1]{K.V.~Golubkov}
\author[2]{N.S.~Gorshkov}
\author[5]{T.I.~Gress}
\author[8]{K.G.~Kebkal}
\author[1]{I.V.~Kharuk}
\author[2]{E.V.~Khramov}
\author[2]{M.M.~Kolbin}
\author[9]{S.O.~Koligaev}
\author[2]{K.V.~Konischev}
\author[2]{A.V.~Korobchenko}
\author[1]{A.P.~Koshechkin}
\author[6]{V.A.~Kozhin}
\author[2]{M.V.~Kruglov}
\author[10]{V.F.~Kulepov}
\author[5]{Y.E.~Lemeshev}
\author[1]{\framebox{M.B.~Milenin}}
\author[5]{R.R.~Mirgazov}
\author[2]{D.V.~Naumov}
\author[6]{A.S.~Nikolaev}
\author[1]{D.P.~Petukhov}
\author[2]{E.N.~Pliskovsky}
\author[11]{M.I.~Rozanov}
\author[5]{E.V.~Ryabov}
\author[1]{G.B.~Safronov}
\author[2,7]{D.~Seitova}
\author[2]{B.A.~Shaybonov}
\author[1]{M.D.~Shelepov}
\author[1]{S.D.~Shilkin}
\author[6]{E.V.~Shirokov}
\author[3,4]{F.~\v{S}imkovic}
\author[2]{A.E.~Sirenko}
\author[6]{A.V.~Skurikhin}
\author[2]{A.G.~Solovjev}
\author[2]{M.N.~Sorokovikov}
\author[4]{I.~\v{S}tekl}
\author[1]{A.P.~Stromakov}
\author[1]{O.V.~Suvorova}
\author[5]{V.A.~Tabolenko}
\author[2]{B.B.~Ulzutuev}
\author[2]{Y.V.~Yablokova}
\author[1]{D.N.~Zaborov}
\author[2]{S.I.~Zavyalov}
\author[2]{D.Y.~Zvezdov}
\author[13]{A.K. Erkenov} 
\author[14,15]{N.A. Kosogorov} 
\author[15]{Y.A. Kovalev} 
\author[15,14,16]{Y.Y.~Kovalev} 
\author[15]{A.V.~Plavin} 
\author[14,15]{A.V.~Popkov}
\author[17,15]{A.B.~Pushkarev} 
\author[18]{D.V.~Semikoz} 
\author[13]{Y.V.~Sotnikova} 
\author[1]{S.V.~Troitsky}

\affiliation[1]{Institute for Nuclear Research, Russian Academy of Sciences, Moscow, 117312, Russia}
\affiliation[2]{Joint Institute for Nuclear Research, Dubna, 141980, Russia}
\affiliation[3]{Comenius University, 81499 Bratislava, Slovakia}
\affiliation[4]{Czech Technical University in Prague, Institute of Experimental and Applied Physics, 11000 Prague, Czech Republic}
\affiliation[5]{Irkutsk State University, Irkutsk, 664003, Russia}
\affiliation[6]{Skobeltsyn Institute of Nuclear Physics, Moscow State University, Moscow, 119991, Russia}
\affiliation[7]{Institute of Nuclear Physics of the Ministry of Energy of the Republic of Kazakhstan, 050032, Almaty, Kazakhstan}
\affiliation[8]{LATENA, St. Petersburg, 199106, Russia}
\affiliation[9]{INFRAD, Dubna, 141981, Russia}
\affiliation[10]{Nizhny Novgorod State Technical University, Nizhny Novgorod, 603950, Russia}
\affiliation[11]{St.~Petersburg State Marine Technical University, St.~Petersburg, 190008, Russia}
\affiliation[12]{Moscow, free researcher}
\affiliation[13]{Special Astrophysical Observatory of RAS, Nizhny Arkhyz 369167, Russia}
\affiliation[14]{Moscow Institute of Physics and Technology, Institutsky per. 9, Dolgoprudny 141700, Russia}
\affiliation[15]{Astro Space Center of Lebedev Physical Institute, Profsoyuznaya 84/32, 117997 Moscow, Russia}
\affiliation[16]{Max-Planck-Institut für Radioastronomie, Auf dem Hugel 69, 53121 Bonn, Germany}
\affiliation[17]{Crimean Astrophysical Observatory, Nauchny 298409, Crimea, Russia}
\affiliation[18]{APC, Universite Paris Diderot, CNRS/IN2P3, CEA/IRFU, Sorbonne Paris Cite, 119 75205 Paris, France}

\abstract{We report on the observation of a rare neutrino event detected by Baikal-GVD in April 2021. The event GVD210418CA is the highest-energy cascade observed by Baikal-GVD so far from the direction below the horizon. The estimated cascade energy is $224\pm75$~TeV. The evaluated signalness parameter of GVD210418CA is 97.1\% using an assumption of the E$^{-2.46}$ spectrum of astrophysical neutrinos. The arrival direction of GVD210418CA is near the position of the well-known radio blazar TXS~0506+056, with the angular distance being within a 90\% directional uncertainty region of the Baikal-GVD measurement. The event was followed by a  radio flare observed by the RATAN-600 radio telescope, further strengthening the case for the neutrino-blazar association.}

\ConferenceLogo{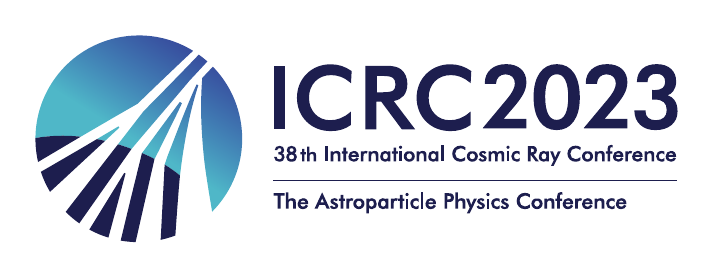}

\FullConference{%
38th International Cosmic Ray Conference (ICRC2023)\\
  26 July-3 August 2023\\
  Nagoya, Japan}

\begin{document}
\maketitle

\section{Introduction}
Baikal-GVD, presently the largest neutrino telescope in the Northern Hemisphere, has achieved the expected sensitivity to astrophysical neutrino fluxes measuring high-energy cascade-like events in the data collected since 2018~\citep{ZhD-Neutrino2022}. The statistical significance of 2.22$\sigma$ was obtained in the all-sky cascade analysis for the energies of above 70 TeV, as well as 3.05$\sigma$ for the data sample containing only upward-going events with the energies of above 15 TeV~\citep{GVD-Diffuse2022}. New data on astrophysical neutrinos observed with Baikal-GVD complement the latest landscape of high-energy and ultra-high-energy neutrinos and the searches for neutrino sources across different energies and distances covered~\citep{Snowmass-2022}. 

The first association of an individual astrophysical object with a high-energy neutrino was found for the blazar TXS~0506+056~\citep{IC170922A}. On September 22, 2017, the IceCube  neutrino obsesrvatory detected a probable muon neutrino with an energy of 290 TeV from the direction of this blazar -- the neutrino alert IC170922A. It was followed by a large flash from TXS~0506+056 observed simultaneously in the entire electromagnetic spectrum -- from radio to  gamma rays~\citep{multi_TXS2018}. Moreover, this event was supplemented by about a 3.5$\sigma$ excess of lower-energy neutrinos from the direction of TXS~0506+056 found by IceCube in their earlier data sample from 2014-2015. Also, the ANTARES neutrino telescope reported on one low-energy muon neutrino event from the direction of the blazar TXS~0506+056 although the event corresponded to the background~\citep{ANTARES:2018osx}. The assumption about extragalactic origin of some astrophysical neutrinos from the objects like active galactic nuclei (AGN) and, in particular, from blazars (objects of an AGN class with relativistic jets directed towards the Earth) was made quite a long time ago~\citep{Berezinsky-1997}, even before the first data from a large-scale neutrino telescope were obtained. Recent studies of spatial coincidence between high-energy neutrinos observed and bright radio blazars~\citep{CT_2020_1} have updated existing ideas. In summary, part of high-energy neutrinos is likely to be associated with radio-loud blazars whose population in the diffuse flux does not exceed 30\%~\citep{Murase-2018}. Some questions remain open about specific blazar areas where neutrinos are produced and about physical connection between outgoing neutrinos and electromagnetic flares.

The Baikal-GVD neutrino experiment is aimed at detecting neutrino signals from astrophysical sources while the telescope is still under construction. In the sample of the cascades selected for 4 years, 25 high-energy events have been detected~\citep{ZhD-Neutrino2022}. Half of them may have astrophysical origin~\citep{GVD-Diffuse2022}. The observation of high-energy neutrinos with the energies of above 200 TeV definitely indicates the acceleration of cosmic particles far away from the Earth in our Galaxy and beyond. Baikal-GVD has detected such an event with an energy estimated to be $224\pm75$~TeV in the data sample containing only upward-going events with energies of above 15 TeV. The cascade with the highest energy, detected by Baikal-GVD below the horizon so far, arrived from the direction of the well-known blazar TXS~0506+056~\citep{GVD-TXS2022}. The blazar TXS~0506+056 is characterized by a bright compact parsec-scale emission, but appears to be otherwise completely ordinary~\citep{Kovalev-2020}. We estimated the probability of the cascade event with an energy of above 200 TeV, induced by a neutrino inside the GVD volume, of having astrophysical origin and the likelihood of its association with TXS~0506+056. We also investigated the electromagnetic activity of the blazar.
\\
\section{Baikal-GVD neutrino experiment}
Baikal-GVD is a deep under-water neutrino telescope of gigaton volume in Lake Baikal. Every year, since 2016, the effective volume of the Baikal-GVD telescope increases - one or two clusters of strings with optical modules are deployed in Lake Baikal at the depths between 0.75 and 1.3 km. In 2022, the facility volume exceeded 0.4 km$^3$ for the energy range of above 100 TeV. In 2023, a total of 96 strings with 3456 optical modules were installed. From the earliest stages of construction, Baikal-GVD has been accumulating the data samples of neutrino candidate events by measuring Cherenkov radiation from interactions of high-energy neutrinos in water. Neutrino-nucleon interactions produce both cascade-like events from generated electro-magnetic and hadronic showers and also track-like events in case of muon production. Both types of events are reconstructed with the software~\citep{GVD-BARS2018} developed by Baikal-GVD analysis groups. The current design and status of the Baikal-GVD detector are presented at this conference (PoS(ICRC2023)976).
%(NU1-03). 

\begin{figure}

    \centering
   \includegraphics[trim=0cm 0cm 0cm 0cm,width=0.45\linewidth,height=275px]{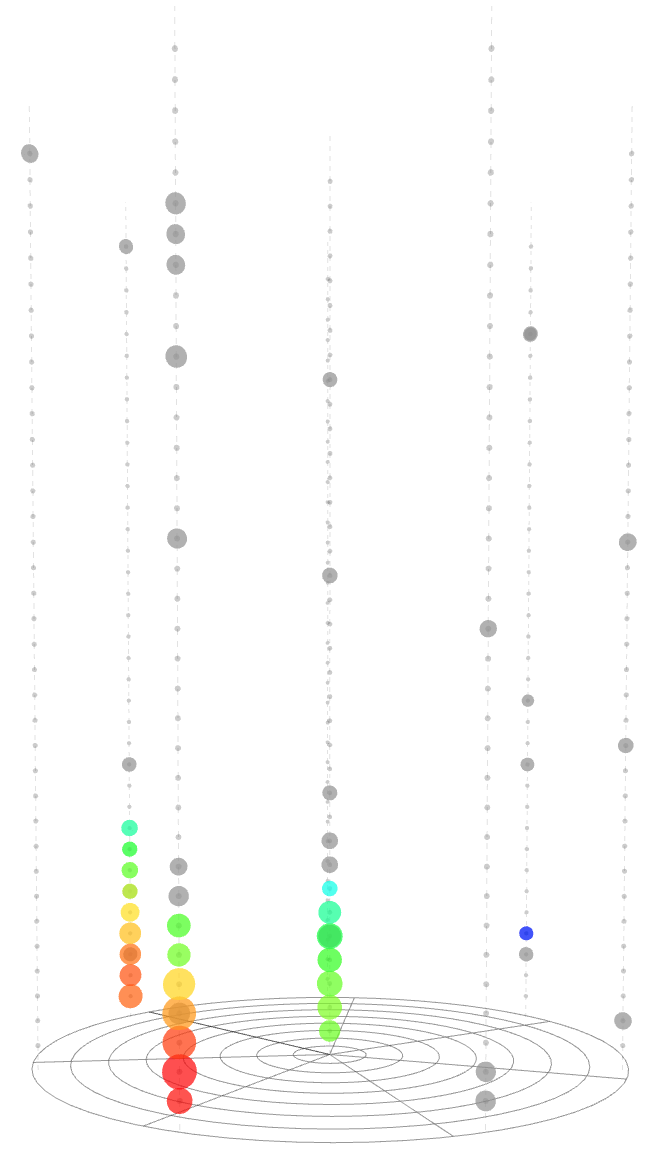} 
   \includegraphics[trim=0cm 0cm 0cm 0cm,width=0.35\linewidth,height=275px]{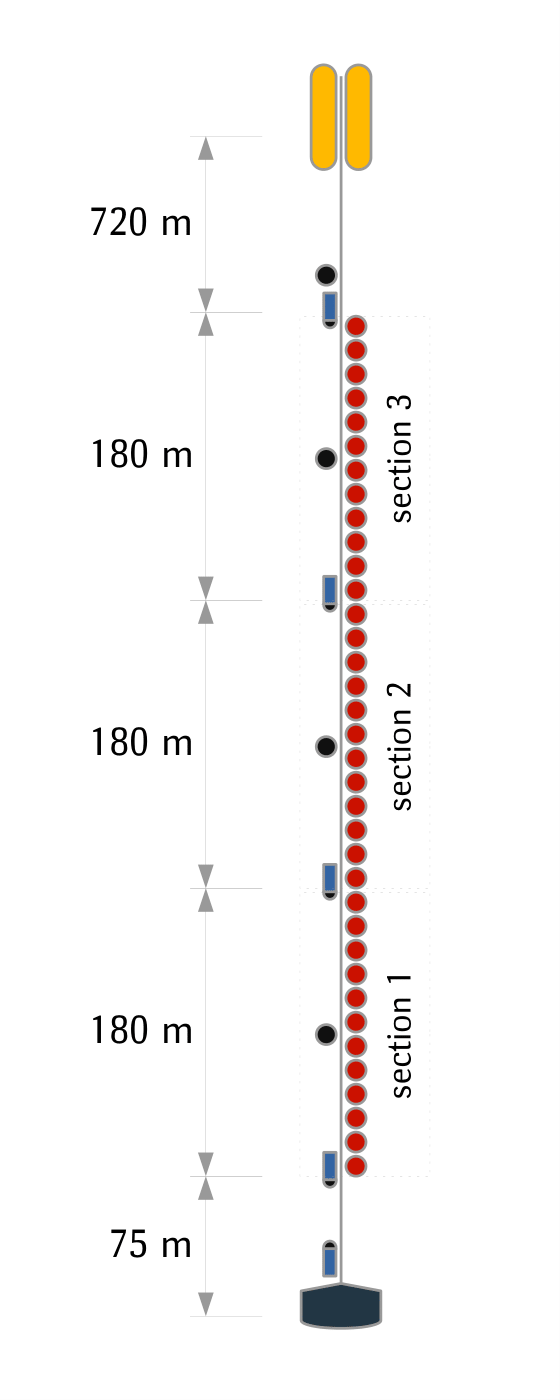}
    \caption{Left: response of the Baikal-GVD detector to the GVD210418CA event.
    The colour gradient corresponds to the earlier (red) and later (blue) photon arrival time in hit optical modules while the size of spheres is a measure for the recorded number of 
     photoelectrons. Grey points represent hits excluded from the analysis (noise hits). Right: equipment placed on a string %.Colour designation of circles are: 
     where OMs are red circles, the string master module (top) and three-section master modules are in black, the acoustic modems are in dark blue. %Note that a string has only one of the two lowest AMs shown in the scheme. If an AM is placed near the anchor (black box on the bottom) it works as an emitter. 
     The buoy (yellow) is placed on the top at a depth of 30 m under the lake surface.
    }
    \label{f:clsTXS}
\end{figure}

Basic elements of Baikal-GVD are optical modules (OMs). Inside pressure-resistant glass spheres of OMs, there are  photomultipliers Hamamatsu R7081-100 (PMTs) with a photocathode 10 inches in diameter and electronics. OMs are mounted onto vertical cables forming a string. Each string comprises 36 OMs spaced vertically $\sim15$ m apart at a depth of 750~m to 1275~m as shown in \autoref{f:clsTXS}. Strings with OMs are collected in clusters. Each cluster is an independent array comprising 8 strings with a total of 288 OMs and is connected to the Shore Station by its own electro-optical cable. Seven of these eight strings are arranged in a heptagonal grid with $\sim60$~m spacing. Inter-cluster distances between the central strings vary from 250 to 300~m. 
In 2023, Baikal-GVD has 12 clusters consisting of 3456 photodetectors and systems for recording, collecting and transmitting data to the shore and processing them online.

In 2019, the Baikal-GVD experiment reported on the first few  $E>$100 TeV neutrino candidates~\citep{GVD-2019} that are complementary to the IceCube alerts for neutrinos incoming from the Northern Hemisphere. Recently, Baikal-GVD has given the first independent confirmation to the visible astrophysical component in measured diffuse neutrino fluxes with the data sample of cascades collected between April 2018 and March 2022~\citep{GVD-Diffuse2022}. In the present study, our greatest interest is directed to the selection of astrophysical candidate neutrinos of the highest purity, containing only upward-going cascade events with zenith angles satisfying $\cos\theta<-0.25$, and also the Baikal-GVD cascade with the highest energy $E>$200~TeV.

\section{Search for high-energy neutrino-induced cascades with single Baikal-GVD clusters}
The analysis is based on the same selection of events as in~\citep{GVD-Diffuse2022} with showers generated in the sensitive volume of the Baikal-GVD neutrino telescope. Shower-like event selection and cascade reconstruction using the data from a single cluster of Baikal-GVD are presented elsewhere, e.g. in~\citep{Baikal-JETP2022}. The Cherenkov radiation from electromagnetic and hadronic showers is formed by photons emitted by charged particles of the shower, mainly by electrons and positrons, and is determined by their spatial, angular and temporal distributions. The cascades are reconstructed as point-like sources of light. The data analysis includes Monte Carlo (MC) production of cascade events in the energy range from 1 TeV to 10 PeV and the OM response function to Cherenkov light as a function of the distance to the OM and photon arrival direction and time. So far, these parameters compose a multi-dimensional array of the Cherenkov cascade photon field for reconstruction of events. Optical properties of the Baikal water at different depths are accounted (PoS(ICRC2023)977), as well as the OM angular acceptance~\citep{GVD-OM2016}. The shower coordinates are estimated by means of a $\chi^2$ fit over the temporal information of the telescope's triggered channels. After that, the shower energy and angular directions are reconstructed by the maximum likelihood method using the shower vertex coordinates, the amplitude information from OMs and the characteristic angular distribution of the Cherenkov cascade photon field. The values of the variables $\theta$, $\phi$ and $E_{sh}$ corresponding to the maximum value of the optimization functional are chosen as the polar and azimuthal angles characterizing 
the direction of the development of the shower and its energy. 

 The precision of the energy reconstruction is $\sim(10-30)\%$ and depends substantially on the energy of the cascade, its position and orientation relative to the cluster. The precision of reconstruction of the progenitor neutrino direction depends strongly on OM hit multiplicity and is about $2^\circ - 4^\circ$ (median value). In the data collected between April 2018 and March 2022, the rate of triggered events is about 100 Hz while equivalent live time for a single cluster corresponds to 13.5 years.  After  noise-hit suppression, cascade reconstruction, application of cuts to reconstruction quality parameters, a sample of 14328 cascades with the reconstructed energy $E_\mathrm{sh}>10$~TeV and OM hit multiplicity $N_\mathrm{hit}>$11 was selected. Most of these events have atmospheric origin and constitute the background for the astrophysical neutrino search. Additionally, a drastic background suppression is achieved by selecting only upward-moving cascades as it was described in~\citep{GVD-Diffuse2022}. The cascade-like events with $E_\mathrm{sh}>15$~TeV and reconstructed zenith angle $\theta$ satisfying $\cos\theta < -0.25$ were selected as astrophysical neutrino candidates. A total of 11 such events have been found in the data sample while, on average, 2.7 events are expected from atmospheric neutrinos and 0.5 events from misreconstructed atmospheric muons. The most energetic cascade among those 11 events was reconstructed with the energy $E_\mathrm{sh}$ exceeding 200 TeV and with the number of hits $N_\mathrm{hit} = 24$ while in other cascades the estimates of energies were less than 100 TeV. This event is further discussed in the next section.

\section{A high energy Baikal-GVD neutrino from the direction of TXS~0506+056}
 
\begin{figure}
    \centering
    \includegraphics[trim=0cm 0cm 0cm 0cm,width=0.42\linewidth,height=130px]{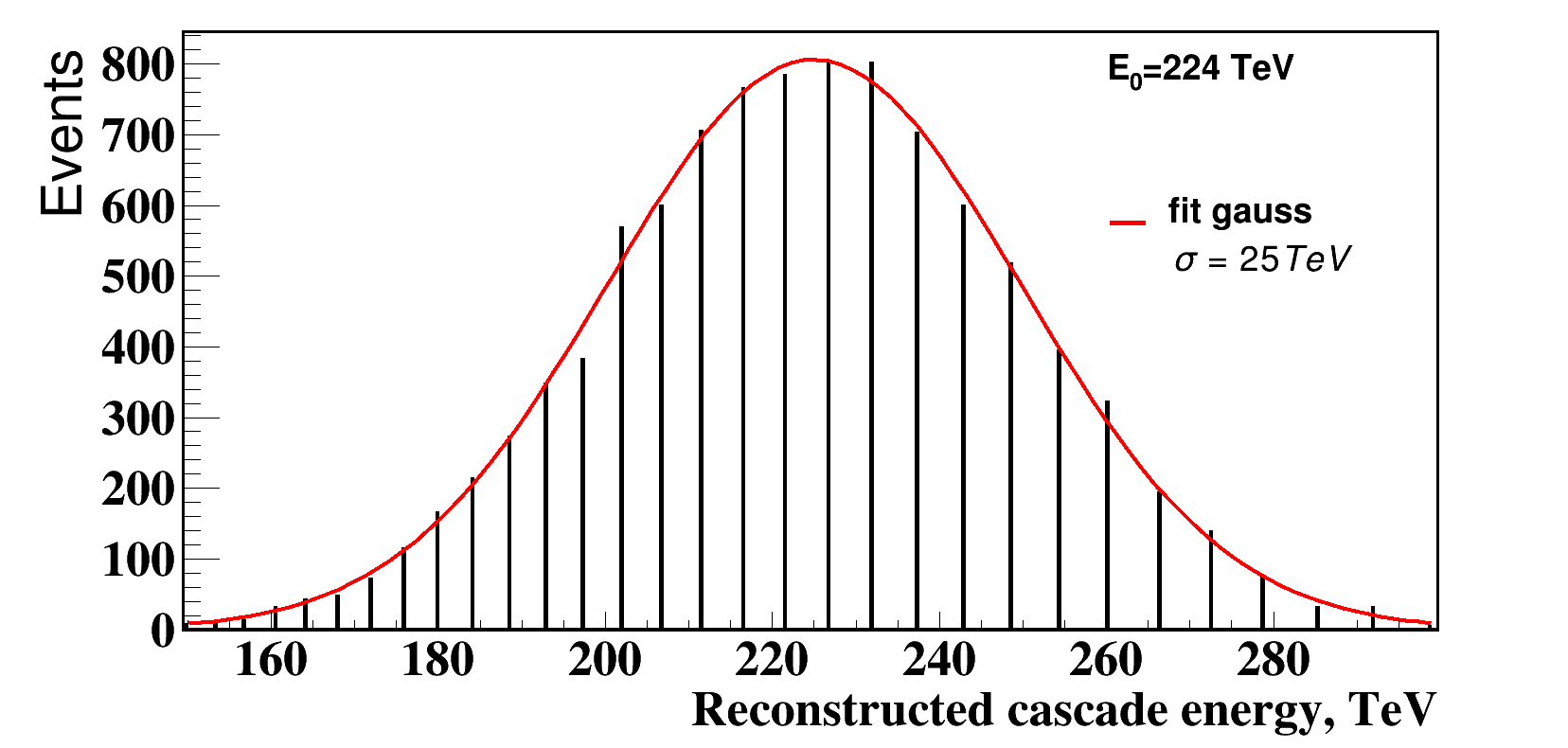}
    \includegraphics[trim=0cm 0cm 0cm 0cm,width=0.57\linewidth,height=130px]{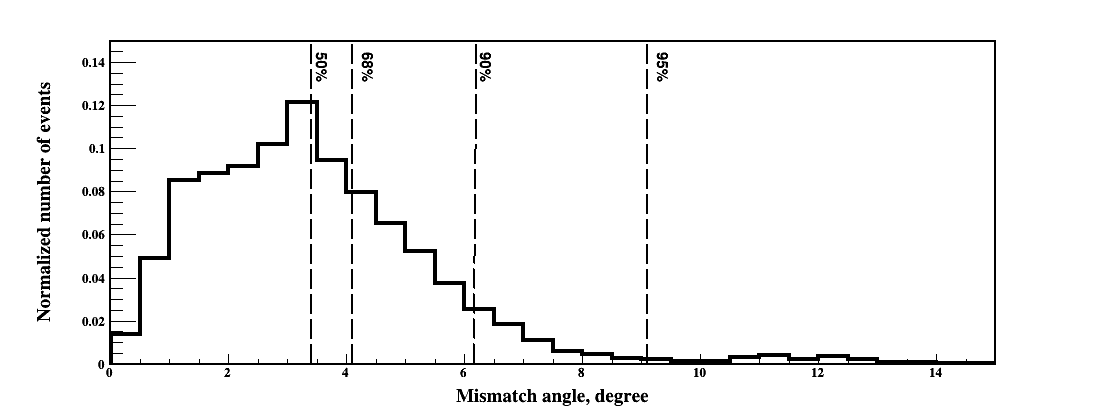}
    \caption{
    %\textit{Left:} 
    Left: distribution of the reconstructed energy of Monte Carlo cascades with a fixed energy of 224~TeV of the generated cascade; the fitted Gaussian $\sigma=25$~TeV.
    %\textit{Right:} 
    Right: distribution of Monte Carlo simulated differences between generated and reconstructed arrival directions. The 50\,\% mismatch angle is $3.4^\circ$, for 90\,\% we found $6.2^\circ$.
    This was obtained taking into account the uncertainty in the recovered energy of the event.
    \label{f:SB}}
\end{figure}

\begin{figure}
    \centering
    \includegraphics[width=0.45\linewidth,height=130px]{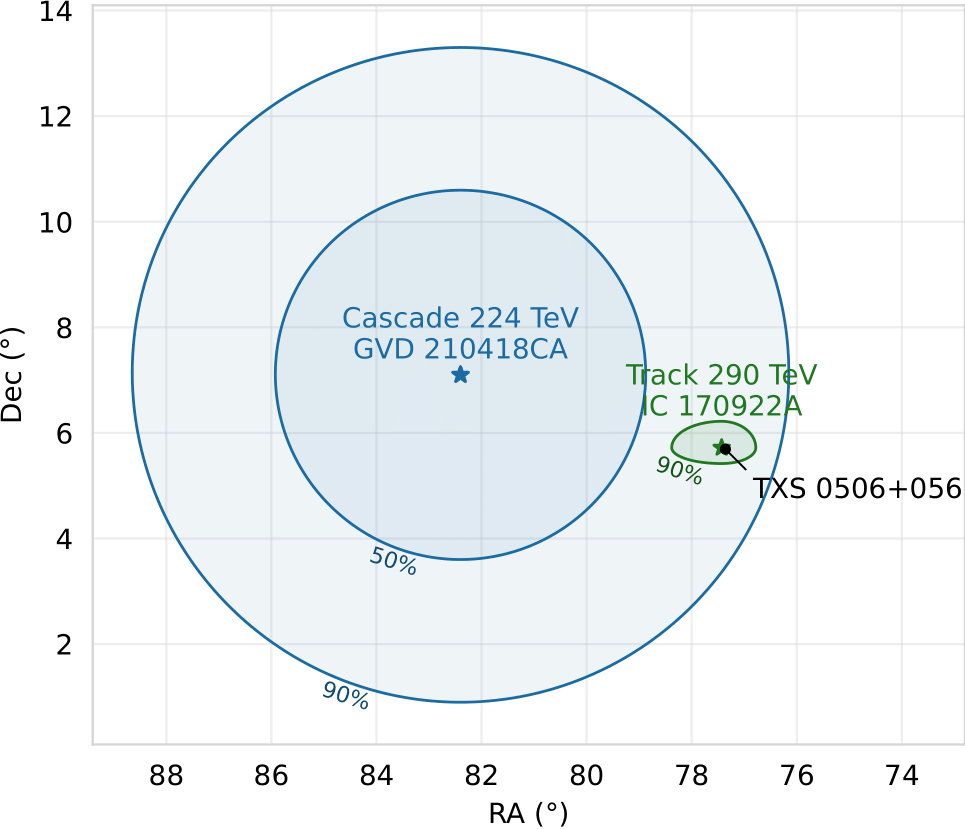}
    \caption{Arrival directions of the 2017 IceCube track event and the 2021 Baikal-GVD cascade event with their directional uncertainties, together with the position of TXS~0506+056.}
    \label{f:mapTXS}
\end{figure}

The highest-energy upward-going cascade was detected by Baikal-GVD on April 18, 2021, at the arrival time 22:45:55 UTC~\citep{GVD-Diffuse2022, GVD-TXS2022} and was named GVD210418CA. The pattern of this cascade event detected by
one cluster is shown in \autoref{f:clsTXS} on the left panel while on the right panel all structure elements of a string of an eight-string cluster are shown. The value of the zenith angle of the GVD210418CA cascade is 115$^\circ$. The precision of reconstruction of the arrival direction of the neutrino which caused the GVD210418CA cascade was estimated by reconstruction of Monte Carlo generated cascades with energies and directions equal to reconstructed values of the event. This takes into account the uncertainty in energy reconstruction. The distribution of the reconstructed energy is shown in \autoref{f:SB} on the left panel. At a three-sigma confidence level, the energy error is 75 TeV. We use Monte Carlo simulations of isotropically distributed incoming astrophysical neutrinos of the observed energy (224~TeV) with respect to detector exposure which is not uniform and depends on the zenith angle. The Earth is not fully transparent to neutrinos at the highest energies. And for neutrinos with the energy of about 200 TeV and $\theta\sim115^\circ$, the not-full-transparency is negligible. The distributions of the mismatch angle between generated and reconstructed directions is shown in \autoref{f:SB} on the right panel. Monte Carlo simulations of this event demonstrate that in 50\% of cases the reconstructed neutrino arrival direction is within the opening angle of $3.2^\circ$ from the simulated one, and in 90\% of cases the angle is within 6.2$^\circ$. The equatorial coordinates of the cascade GVD210418CA at observation time are  RA$= 82.4^\circ$ and DEC$= +7.1^\circ$ as shown in \autoref{f:mapTXS} with 50\% and  90\% containment circles. This 90\% containment circle contains the direction to the TXS~0506+056 blazar as illustrated in \autoref{f:mapTXS}. 

Using Monte Carlo samples of cascades in the energy range from 10 TeV to 10 PeV and flux normalizations for incoming  muons and neutrinos, conventional and prompt, as in~\citep{GVD-Diffuse2022}, we determine the probability of the event to be the background from atmospheric processes for the given $E_\mathrm{sh}$,  $N_\mathrm{hit}$, and shower direction. The atmospheric origin of the event is found to be as low as $P_\mathrm{atm} = 0.0033$ for the background-only hypothesis. Further, assuming the $E^{-2.46}$ spectrum of astrophysical neutrinos~\citep{IceCube:HESE2020}, we determine the ratios of the expected numbers of the signal to the sum of background and %signal events in dependence on cascade energies and cosine zenith angles (see poster PMM 1-10). The parameter of ratios called signalness was introduced by IceCube 
signal events in dependence on cascade energies and cosine zenith angles (PoS(ICRC2023)1458). The parameter of ratios called signalness was introduced by IceCube earlier. For given $E_\mathrm{sh}$ and $\theta_\mathrm{sh}$, the signalness of the GVD210418CA cascade is 97.1\%. For comparison, the signalness of the %the IceCube earlier~\citep{IceCubeOldAlerts}. For given $E_\mathrm{sh}$ and $\theta_\mathrm{sh}$ the signalness of the GVD210418CA cascade is 97.1\%. For comparison, the signalness of the
track event IC-170922A, associated with TXS~0506+056, was 56.5\% ~\citep{multi_TXS2018}. 

A significant correlation between the neutrino arrival and the gamma-ray activity of a source was found for one case only, namely for the neutrino event claimed by the IceCube Collaboration at the estimated neutrino energies $E=290^{+2010}_{-75}$~TeV~\citep{multi_TXS2018} and a gamma-ray flare TXS 0506+056 on September 22, 2017. Thus, we have found an association of the exceptional Baikal-GVD event, GVD210418CA, having a 99.67\% probability to be of astrophysical origin, with the exceptional source singled out by previous studies. The possibility of associating the highest-energy upward-going cascade GVD210418CA with this well-known blazar is considered by Baikal-GVD in~\citep{GVD-TXS2022} and discussed below.

\section{Discussions}
\begin{figure*}
    \centering
    \includegraphics[width=0.9\linewidth,height=130px]{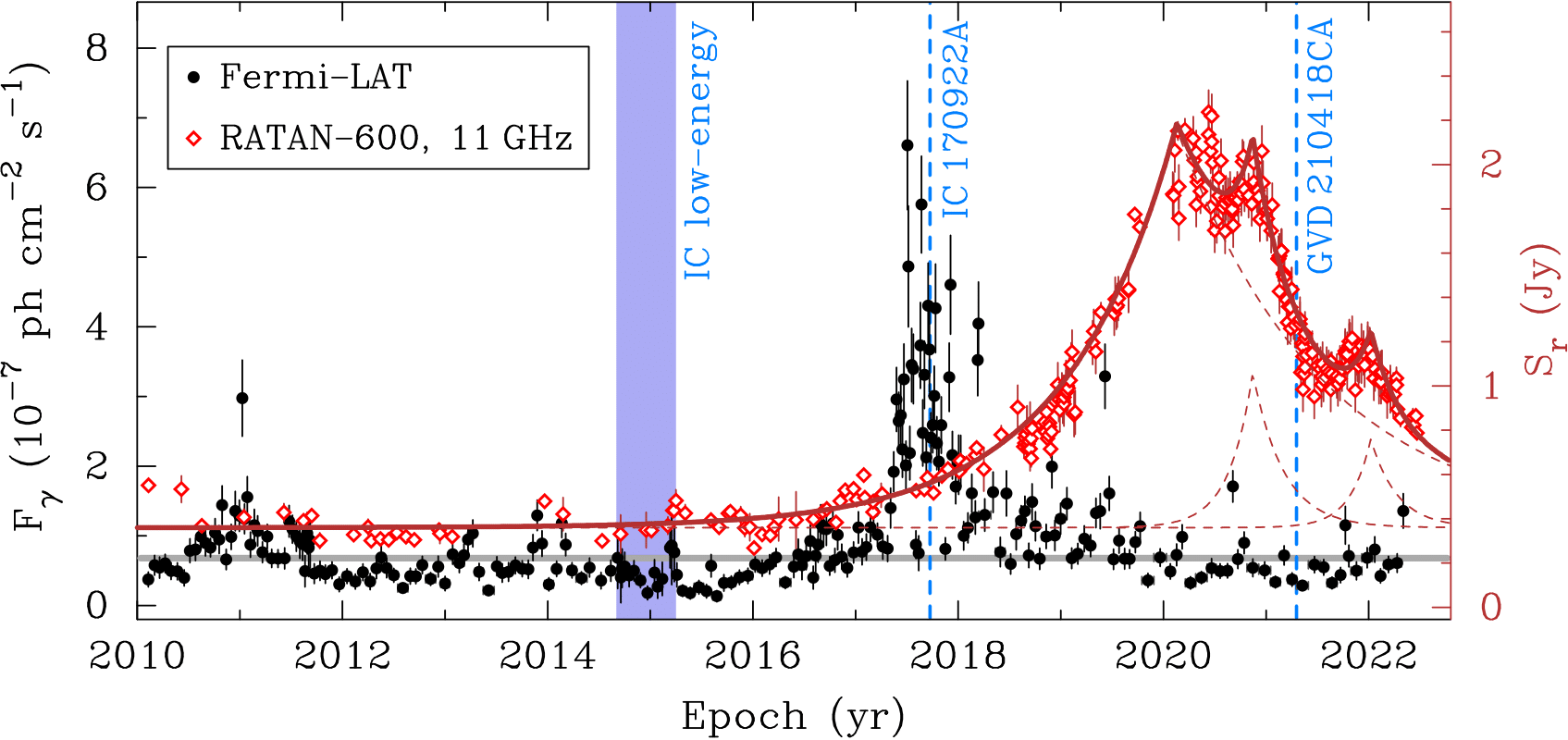}
    \caption{
    Radio and gamma-ray light curves of TXS~0506+056 (all details of flare decompositions are presented in~\citep{GVD-TXS2022} and references therein). Black dots: \textit{Fermi} LAT adaptive binning light curve of the gamma-ray source 4FGLJ0509.4+0542 positionally associated with  TXS~0506+056. The grey horizontal line indicates the median gamma-ray flux. Empty red diamonds: RATAN-600 light curve at 11~GHz. The radio light curve is decomposed by three radio flares depicted by dark-red dashed lines, the sum of which is represented by a thick line.}
    \label{f:neutrino_Fermi_radio_lc}
\end{figure*}

%%{The 600-meter ring radio telescope RATAN-600 of the Special Astrophysical Observatory, the Russian Academy of Sciences, performs long-term observations of continuum spectra at centimeter wavelengths of a large sample of extragalactic radio sources selected with very long baseline interferometry (VLBI) technique~\citep{Kovalev}}
   
All extremely high-energy events, which used to indicate TXS 0506+056 as a neutrino source, have $E>200$~TeV, and this threshold was previously chosen for various other analyses \citep[e.g.~by][]{IceCube-2016}. In what follows, we also adopt this energy threshold and notice that lowering the threshold to $\sim$100~TeV would not change our further numerical estimates considerably. 

The highest-purity sample of astrophysical neutrino candidates in the Baikal-GVD analysis contains upward-going cascade events only, with the cut on zenith angles satisfying $\cos\theta<-0.25$. Only one Baikal-GVD cascade with $E>200$~TeV was detected from this $0.75 \cdot 2\pi$~sr of the sky. The chance probability that the arrival direction of a single event coincides with the position of TXS~0506+056 in the sky is found to be less than 0.8\%  within a 90\,\% confidence level angular resolution cone of $6.2^\circ$ (see \autoref{f:SB}, on the right). 

In 2021, at the time of the Baikal-GVD event GVD210418CA, eight clusters of Baikal-GVD were in operation, with the effective area for cascades several times smaller than that of IceCube for the track channel. The observation of one event in Baikal-GVD and zero events in a larger IceCube observatory are reconcilable with the Poisson fluctuations provided that the expected number of the observed events is small. This holds only under the assumption of a flaring source, with the duration of flares about a year or shorter. Otherwise, non-observation of events by IceCube\footnote[1]{On September 18, 2022, IceCube sent an alert for the neutrino with an energy of $\sim$170 TeV arrived in directional coincidence with the blazar TXS~0506+056. Baikal-GVD followed up this alert and presented results in PoS(ICRC2023)1458.} would become incompatible with the assumed flux due to its large cumulative exposure.   

%The estimated neutrino flux\footnote[2]{Assumptions and spectra E$^{-2}$ and E$^{-2.5}$, as in~\citep{GVD-TXS2022}.} corresponds to $E^2F(E) \sim 10^{-11}$~TeV\,cm$^{-2}$\,s$^{-1}$, 
The estimated neutrino flux (in assumptions of spectra E$^{-2}$ or E$^{-2.5}$ as in~\citep{GVD-TXS2022}) corresponds to $E^2F(E) \sim 10^{-11}$~TeV\,cm$^{-2}$\,s$^{-1}$, consistent by an order of magnitude with the IceCube estimates of 2017 and with the gamma-ray flux in the \textit{Fermi} LAT band during the high state of TXS~0506+056 \citep{multi_TXS2018}. However, the \textit{Fermi} LAT flux remains low in the period of detection of the Baikal-GVD event. The earlier observation of the lack of gamma-ray activity during the neutrino flare of the same source in 2014 (\autoref{f:neutrino_Fermi_radio_lc}) suggests that simple one-zone models are disfavored as an explanation of both gamma-ray and neutrino data. For the recent discussion see~\citep{CT-UFN2021}. However, \autoref{f:neutrino_Fermi_radio_lc} reveals an important similarity between the IceCube (2017) and Baikal-GVD (2021) events in terms of radio observations: both neutrinos arrived at the beginning of remarkable radio flares. Also the Baikal neutrino event detected on April 18, 2021, may coincide with the beginning of a new radio flare peaking at the outset of 2022. Flare decomposition of radio light curves was obtained via observations at the RATAN-600 radio telescope using a long-base interferometer and long-term monitoring of extragalactic radio sources (see~\citep{Kovalev-2020}). Referring to the detailed comparative analysis of the data observed at different wavelengths discussed in~\citep{GVD-TXS2022}, we notice that the number of radio blazars, and TXS~0506+056 among them, may be singled out by prolonged periods of higher activity or just demonstrate accidental fluctuations. It is complicated to correctly estimate a formal probability of chance coincidence, especially due to its \textit{a posteriori} nature. Future data in studies with higher exposures may clarify the origin of events with an energy of at least two hundred TeV in the direction of the TXS~0506+056 blazar.

%\section*{Acknowledgements}
%The research results were obtained using the material and technical base of the Baikal 
%Neutrino Observatory and experimental data accumulated by the Deep Underwater Neutrino
%Telescope Baikal-GVD. 
This work is partially supported by the European Regional Development
Fund -- Project "Engineering applications of microworld physics" (CZ02.1.01/0.0/0.0/16 019/0000766).

%\end{document}

%% Full authors list (ONLY FOR COLLABORATIONS)
%\clearpage
%\section*{Full Authors List: \Coll\ Baikal-GVD Collaboration + expand}
%
%\noindent \textbf{Note comment afterwards:} Collaborations have the possibility to provide an authors list in xml format which will be used while generating the DOI entries making the full authors list searchable in databases like Inspire HEP. \\
%
%\scriptsize
%\noindent
%first.author$^1$, 
%second.author$^2$, 
%third.author$^3$ % .... more names
%and 
%last.author$^{n}$ \\
%
%\noindent
%$^1$first.affiliation.
%$^2$second.affiliation. % .... more affiliation
%$^{m}$last.affiliation.
%\noindent

\end{document}